\documentclass[usenatbib,usegraphicx,letterpaper]{mn2e}

\usepackage{amsmath}
\usepackage{natbib}
\usepackage{color}
\usepackage{soul}
\usepackage{amsmath}
\usepackage{amssymb}

\newcommand{\plotone}[1]{%
 \includegraphics[width=84mm]{#1}%
}%

\newcommand{\msun}{{M}_{\odot}}

\newcommand{\Teff}{T_{\mathrm{eff}}}

\newcommand{\Tin}{T_{\mathrm{in}}}
\newcommand{\rin}{r_{\mathrm{in}}}
\newcommand{\rout}{r_{\mathrm{out}}}
\newcommand{\Lirr}{L_{\mathrm{irr}}}
\newcommand{\Mdot}{\dot{M}}
\newcommand{\Porb}{P_{\mathrm{orb}}}

\newcommand{\RLo}{R_{L,1}}
\newcommand{\ELC}{\textsc{ELC }}

\newcommand{\Lgmin}{L_{g',\phi=0}}
\newcommand{\chig}{\chi^2_{g'}}
\newcommand{\chii}{\chi^2_{i'}}
\newcommand{\ibin}{i_{\mathrm{bin}}}
\newcommand{\delc}{d_\mathrm{\ELC}}

\newcommand{\ees}[2]{\ensuremath{#1 \times 10^{#2}}}

\newcommand{\aj}{{AJ}}%
\newcommand{\araa}{{ARA\&A}}%
\newcommand{\apj}{{ApJ}}%
\newcommand{\apjl}{{ApJ}}%
\newcommand{\aap}{{A\&A}}%
\newcommand{\aaps}{{A\&AS}}%
\newcommand{\mnras}{{MNRAS}}%
\newcommand{\pasp}{{PASP}}%
\newcommand{\nat}{{Nature}}%
\newcommand{\iaucirc}{{IAU~Circ.}}%
\newcommand{\physrep}{{Phys.~Rep.}}%
\title[Optical Light Curve Modelling of SAX J1808]{Optical Observations of SAX J1808.4-3658 During Quiescence}  
\author[Deloye et al.]{C.~J. Deloye$^1$\thanks{E-mail: cjdeloye@northwestern.edu}, C.~O. Heinke$^{1,2,3}$, R.~E. Taam$^{1,4}$, P. G. Jonker$^{5,6}$ \\
$^1$ Northwestern University, Dept. of Physics \& Astronomy, 2145 Sheridan Rd., Evanston, IL 60208 \\
$^2$University of Virginia, Dept. of Astronomy, PO Box 400325, Charlottesville, VA 22902\\
$^3$University of Alberta, Dept. of Physics, 11322-89 Avenue, Edmonton, AB Canada, T6G 2G7\\
$^4$ASIAA/National Tsing Hua University - TIARA, Hsinchu, Taiwan\\
$^5$SRON, Netherlands Institute for Space Research, 3584 CA Utrecht, The Netherlands \\
$^6$Harvard--Smithsonian  Center for Astrophysics, 60 Garden Street, Cambridge, MA~02138
}

\begin{document}

\maketitle

\begin{abstract}
We observed the accreting millisecond pulsar SAX J1808.4-3658 with Gemini-South in $g'$ and $i'$ 
bands, nearly simultaneous with XMM-Newton observations.  A clear periodic flux modulation on 
the system's orbital period is present, consistent with the varying aspect of 
the donor star's heated face. We model the contributions of a disk and donor star 
to these optical bands.  To produce the observed modulation amplitudes, we conclude that the donor must be irradiated by an external flux 2 orders of magnitude greater than provided by the measured X-ray luminosity. A possible explanation for this irradiation is that the radio pulsar mechanism becomes active during the quiescent state as suggested by Burderi et al., with relativistic particles heating the donor's day-side face. Our modelling constrains the binary inclination to be 36--67$^\circ$.  We obtain estimates for the pulsar mass of $>2.2 \msun$ (although this limit is sensitive to the source's distance), consistent with the accelerated NS cooling in this system indicated by X-ray observations.  We also estimate the donor mass to be $0.07$--$0.11 \msun$, providing further indications that the system underwent non-standard binary evolution to reach its current state.
\end{abstract}


\section{Introduction}\label{s:intro}
The low-mass X-ray binary (LMXB) SAX J1808.4-3658 (hereafter J1808) is the first detected accretion-powered millisecond pulsar \citep{chakrabarty98,wijnands98}.  Such objects are thought to be evolutionary intermediaries between LMXBs and radio millisecond pulsars (MSPs): they are still in an LMXB phase with ongoing mass transfer  but have had time for mass transfer to spin up the accreting neutron star (NS) to millisecond spin periods. If and when accretion halts in these systems, they are expected to turn-on as radio MSPs \citep{alpar82,radhak82}.  Since J1808's discovery, seven more accretion-powered millisecond pulsars have been discovered, but J1808 stands out as the best observed member of this class.

J1808 has offered us more than several surprises.  Its NS's mass function \citep{chakrabarty98} implies that the donor companion \citep[which is almost certainly a low-mass brown dwarf][]{bildsten01} has a mass $> 0.04 \msun$ (assuming a fiducial NS mass of 1.4 $\msun$), too large to be explained by standard binary evolution scenarios \citep[see, e.g., figure 5 of][]{deloye08}.  The system's orbital period, $\Porb$, appears to be evolving an order of magnitude faster than binary theory predicts \citep[although, see their discussion of possible explanations for this]{disalvo08,hartman08}.  X-ray observations of the system during its quiescent phases are only able to place upper limits on the NS's thermal component, indicating the NS cools extremely rapidly as compared to most other NSs \citep[see][and references therein]{heinke07}.  Perhaps most surprisingly, observations indirectly suggest J1808 transitions between the LMXB and radio MSP state during X-ray quiescent phases \citep{homer01,burderi03,campana04}.    

The disk in J1808 is thermally unstable and undergoes outbursts lasting approximately a month roughly every 2 years. During the inter-outburst quiescent phase (when the disk is cooler and dim and may not extend inward close to the NS), the system exhibits roughly sinusoidal variability in the optical with the same 2 hr $\Porb$ detected in X-rays \citep{homer01,campana04}.  The phasing of optical maxima corresponds to when the donor is directly behind the NS \citep{homer01}, indicating that the donor's face is heated by flux originating near the NS.  Typically this flux would be due to the X-ray radiation produced by the accretion flow through the inner accretion disk and onto the NS surface.  However, the amplitude of the optical modulation requires significantly greater X-ray flux than is observed during quiescence (albeit the X-ray observations are non-simultaneous with the optical).  This led to speculation  that the NS commences radio pulsar activity during quiescence \citep{burderi03,campana04}, providing the necessary amount of flux to heat the donor's day-side face. 

To investigate J1808 in both optical and X-ray wavelengths nearly simultaneously, 
we have obtained imaging optical data ($g'$ and $i'$) from the Gemini Observatory separated 
by 6.5 hours from an XMM observation of SAX J1808 in quiescence in 2007. Our science 
goals for these observations included further constraints on the thermal component of the X-ray emission and  its X-ray variability in quiescence, simultaneous determination of the optical modulation to constrain definitively its origin, and determining if optical data allows constraints on either the quiescent disk's structure or on the binary's parameters.  The X-ray observations are described in detail in a companion paper \citep{heinke08}.  Here we discuss the Gemini observations.  In \S \ref{s:obs} we describe the observations and data reduction.  In \S \ref{s:lc_analysis}, we describe the light curve analysis within the framework of a theoretical model.  We then discuss the constraints on this model's parameters obtained from our data in \S \ref{s:parconst}.  In \S \ref{s:discussion} and \ref{s:summary}, we discuss our results and summarize.  The Appendix includes details on how our data and further system details (in particular J1808's distance) constrain our model's parameters.

\section{Observations}\label{s:obs}

We observed 1808 on March 8 and March 10, 2007  with the Gemini 
Multi-Object Spectrograph-South \citep[GMOS-S;][]{Hook04} in imaging mode. The March 10 observation  (program GS-2007A-Q-8, UT 06:47 to 09:55) was nearly simultaneous with the  XMM-Newton 
observation (ObsID 0400230501), which started at 16:24 UT, and continued for 16 hours without 
evidence for strong variability.  We took a time series of 44 $g'$ exposures of 185 s each,
with two 230 s $i'$ exposures at either end to obtain colour information. Atmospheric seeing was generally good, ranging from 0.65\arcsec\ to 0.98\arcsec\ FWHM.  Landolt standard 
star fields PG1047+003 and SA104 were also observed during the same night.  
The March 8 observation, using an identical program, suffered from poorer seeing (FWHM ranging from 1.0 to 1.4\arcsec), but still provided data adequate for our purposes, and thus we analyse it as well.

\subsection{Optical photometry}
We used the calibrated Gemini data products, which are processed with Gemini-specific IRAF 
\footnote{IRAF (Image Reduction and Analysis Facility) is distributed by the National Optical 
Astronomy Observatories, which are operated by Association of Universities for Research in 
Astronomy, Inc., under cooperative agreement with the National Science Foundation.} tasks to 
remove the bias, and flat-field and mosaic the images.  We performed photometry on $g'$ and $i'$ 
reference frames using DAOPHOT \citep{Stetson87} point-spread-function (PSF) fitting photometry, 
calibrated using the standard star fields. 
 
\begin{figure}
\plotone{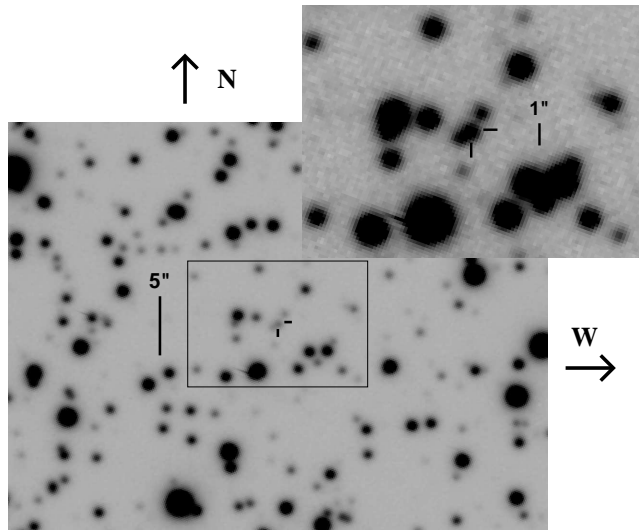}
\caption[fig1.eps]{ \label{fig:fchart}
Finding charts for J1808, average of the seven $g'$ GMOS-S frames with the best (0.65\arcsec) seeing.  A nearby (0.5\arcsec) star can be barely distinguished to the SE.
} 
\end{figure}

Comparison of our images with the finding chart of \citet{homer01} reveals two stars, separated 
by 0.5\arcsec, that are consistent with J1808's indicated location (see Figure \ref{fig:fchart}).
The brighter of these two is variable on a 2-hour period, and we confidently identify it with 
J1808.  By associating ten uncrowded unsaturated nearby stars with stars in the USNO B1.0 catalogue 
\citep{Monet03}, we find a position of $\alpha$=18:08:27.63, $\delta$=-36:58:43.37 (J2000), 
with uncertainties of 0.2\arcsec\ in each coordinate (accounting for the uncertainty in the 
transformation to the USNO B1.0 frame). This is consistent with the VLA-derived position of  \citet{Rupen02},  and with the optical position recently derived by \citet{hartman08}. The optical position derived by \citet{Giles99} in outburst, 1.7\arcsec\ away, is also marginally consistent when the absolute errors in the GSC \citep[up to 1.6\arcsec\ in the southern hemisphere,][]{Taff90}, used as the reference frame by \citet{Giles99}, are considered.

The fainter star ($g'$=22.35$\pm$0.04, $i'$=20.68$\pm$0.02) is located only 0.5\arcsec\ from J1808 
to the SE, at $\alpha$=18:08:27.67, $\delta$=-36:58:43.66.  This star has been identified in 
previous unpublished images of the J1808 field (D. Chakrabarty 2007, priv. comm.), and may 
affect psf-fitting or aperture photometry of J1808 in quiescence taken in poor seeing.  

We then used the differential photometry program ISIS \citep{Alard98,Alard00} for each filter, 
following the basic method of \citet{Mochejska00} as follows.  We select 
the best-seeing reference frame for each filter, and transform the remaining frames onto the 
reference coordinate systems, aligning the frames with bright stars.  We produced a reference 
frame from the seven best-seeing $g'$ frames.  This frame is convolved with a spatially varying 
(3 degrees of freedom) kernel to match each frame's PSF, and subtracted from the remaining 
frames.  Only variable stars (and stars which are saturated) should remain on the subtracted 
images, as stars of constant brightness cancel out.  Finally, profile photometry is extracted 
from the subtracted images, barycenter corrected to TDB using the Perl extension Astro::Time::HJD \footnote{http://search.cpan.org/dist/Astro-Time-HJD/HJD.pm} to correct the observation times to heliocentric Julian dates.  We performed aperture photometry for the standard stars, and used 
this photometry to convert our PSF-fitting photometry, and our differential ISIS photometry, 
into calibrated $g'$ and $i'$ magnitudes. 

\begin{figure}
\plotone{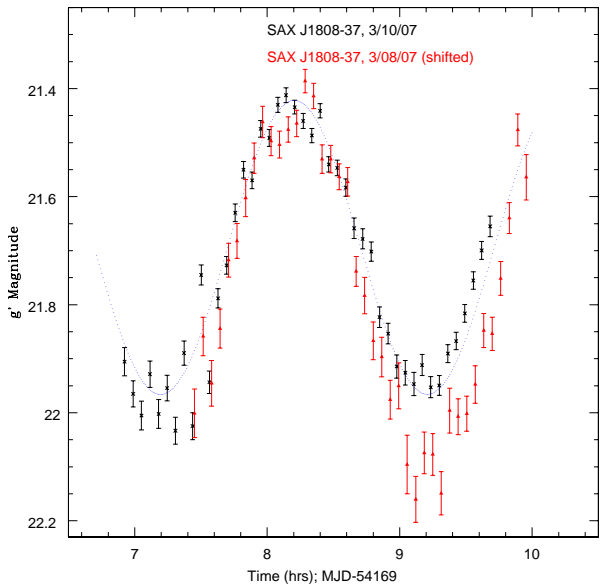}
\caption[fig2.eps]{ \label{fig:glightcurve}
Calibrated $g'$ light curves for J1808, from March 10 (crosses) and March 8 (triangles).  The fit 
shown is to a sinusoid with orbital period fixed to the value of 7249.1569 s \citep{papitto05}.}  
\end{figure}

\section{Analysis of the Optical Light Curves}\label{s:lc_analysis}
\subsection{Data Processing}
The observed optical light curves show significant variability on time scales of order the orbital period as well as flickering on time scales shorter than $\Porb$. Having only obtained four $i'$ values during each night's observations precludes averaging the $i'$ data between nights;  the intrinsic variability of the source further precludes a simple combination of $i'$ data into a single light curve.  We proceed by analysing each night's light curves separately. Later we'll consider what constraints on J1808's system parameters can be derived by considering the two nights' data in tandem.

\begin{figure}
\plotone{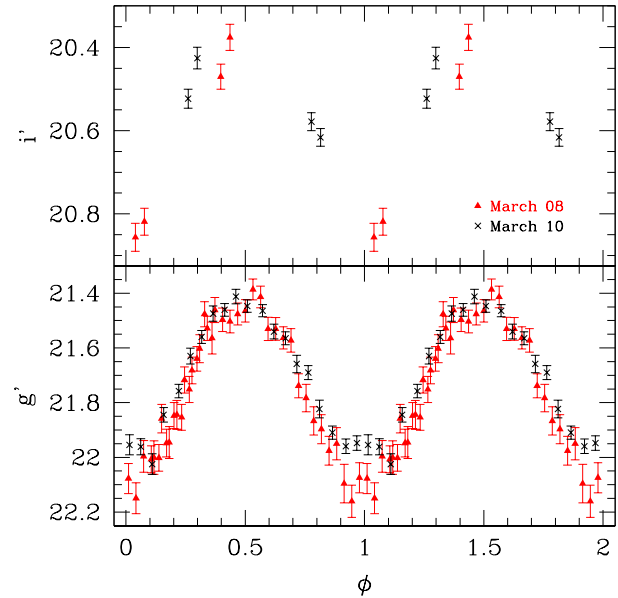}
\caption{The final form of the March 08 and March 10 optical data used in our model light curve 
fits.  Two complete orbits are shown for clarity. }
\label{fig:lc_procdata}
\end{figure}

In order to compare against our model light curves, we sorted our data by orbital phase, $\phi$ using the orbital ephemeris of \citet{papitto05}\footnote{Using the more recent ephemeris of \citet{hartman08} would produce insignificant changes.}.  The March 10 $g'$ data exhibits a clear, systematic brightening beyond the second observed minimum (i.e., past $\approx 9.1$ hrs in figure \ref{fig:glightcurve}), as well as large amplitude flickering. This aperiodic variability is likely due to stochastic processes in the disk (e.g, hot-spot variability) that are not captured by our models described below.  We do not have a sufficient number of observations to average over the stochastic variability. So, in order to compare our models with a observational time-frame representing as much as possible almost constant conditions in the system, we excluded from our analysis the March 10 $g'$-data beyond the second flux minima.  This keeps the $g'$ points obtained during the same orbit as the March 10 $i'$ data.  The remaining $g'$ data we phase sorted, binned (using a bin width $\Delta \phi=0.05$), and then averaged by bin to reduce the flickering's impact on the fits.  The data resulting from all these procedures, and used in our model light curve fitting below, are shown in Figure \ref{fig:lc_procdata}.

\subsection{Theoretical Light Curve Modelling}
We utilize the program \ELC \citep{orosz00} to calculate theoretical model light curves.  To model the optical data, we consider contributions from both an accretion disk and a donor that is irradiated by an energy flux originating at the location of the accreting NS. The \ELC program includes the effects of limb darkening (implicitly via use of \textsc{PHOENIX}\footnote{http://www.hs.uni-hamburg.de/EN/For/ThA/phoenix/index.html} model atmospheres) and disk/donor occultations.  Our model has seven independent input parameters.

The first two are the masses of the accreting NS, $M_1$, and the donor, $M_2$. Once $M_1$, $M_2$ are specified, the binary's inclination relative to our line of sight, $\ibin$ is determined via the NS's mass function \citep{chakrabarty98}. We determine the binary's orbital separation from $M_1$, $M_2$, and $\Porb$. From $\Porb$  and the requirement that the donor fills its Roche lobe, the donor's radius, $R_2$, is determined once $M_2$ is specified. This also fixes the donor's unirradiated effective temperature, $\Teff$, which we obtain from  unpublished low-mass brown dwarf models of Deloye \& Taam \citep[the $\Teff$ obtained from these models are very similar to those of, e.g.,][]{baraffe98}.  The resulting $\Teff$, which are a function of $M_2$ alone, range between $\approx 2000$--$3000$ K. 

To account for the periodic variability, the donor's optical contribution must be phase dependent. The \ELC program allows modelling this source of variability by irradiating the donor with a point source of flux at the accretor's location.  We consider the impact of such a point source with total luminosity, $\Lirr$, between $10^{33}$ and $10^{34.6}$ erg s$^{-1}$.  We fix the donor's albedo in \ELC to the value minimizing the required $\Lirr$ for a given light curve amplitude \footnote{This corresponds to an \ELC input parameter \texttt{alb1} value of 1.0.}. If the donor's albedo differ from this, a commensurate increase in $\Lirr$ will be required to produce the same results.

To model the disk in J1808, we use four parameters: the disk's inner and outer radii, $\rin$ and $\rout$, the temperature of the disk at $\rin$, $\Tin$, and an exponent, $\xi$, that characterizes the disk's radial temperature profile, $T(r) \propto \Tin (r/\rin)^\xi$. The structure of a quiescent accretion disk, in particular is not expected to be that of the standard steady-state disk model \citep[e.g.,][]{dubus01}. This expectation has been verified by observations of other binaries in quiescence, which find, in particular, $\xi$-values ranging from $\xi < -1$ to $\xi \approx 0$  \citep[e.g.,][steady-state disks are expected to have $-0.75 < \xi < -0.5$]{skidmore00,vrielmann02, shabaz03,baptista04}.  Given this expectation, we treat each of these four parameters as independent  quantities.

The values of $\Lirr$ used here are substantially greater than the  $L_X \approx 10^{32}$ ergs s$^{-1}$ observed from 1808 in quiescence \citep{heinke08}. As seen below, such high values of $\Lirr$ are required to fit our data, making it clear that the X-ray emission in the system insufficient to drive the observed optical variability.  As has already been suggested \citep{burderi03,campana04}, the NS itself could provide the required irradiating flux in the form of a pulsar-wind that turns-on during the quiescent phase of the disk outburst cycle.  However, we will refrain from detailed attempts to interpret the source or quality of the required irradiating energy. Instead, we view our results as a guide to the amount of flux required to impinge on the donor in order to produce the observed amplitude of variability. We will make a few further comments on this point in the discussion.

We also attempted to fit the optical data using an irradiated-donor only model.  For this case, instead of fixing the donor's unirradiated $\Teff$ at the values predicted by brown dwarf evolution models, we allowed $\Teff$ to be a  free parameter ranging between $4500$ and $7500$ K (chosen to provide sufficient $g'$ flux at light curve minimum given J1808's distance as discussed below). This provided an input parameter set of $M_1$,  $M_2$, $\Teff$, and $\Lirr$.  This model was motivated by the possibility that horizontal fluid currents in the donor's atmosphere may efficiently advect the irradiating energy to the donor's night side, elevating the surface temperature there \citep[see, e.g.,][]{burkert05,dobbsdixon08}. However, this model can not simultaneously reproduce the light  curves' morphology, amplitude, and $i'-g'$ colours, and did not produce acceptable fits.

We used \ELC to calculate model light curves at discrete values of six of our input parameters: $M_1= $ 1.4, 1.6, 1.8, ..., 3.0 $\msun$; $M_2=$ 0.05, 0.06, 0.07, ..., 0.13 $\msun$; $\rin=$ 0.001, 0.01, 0.03, 0.06, 0.1, and 0.2 $\RLo$ (where $\RLo$ is the Roche lobe radius of the accretor and is the internal unit used by \ELC); $\rout=$  0.25, 0.4, 0.5, 0.6, 0.7, 0.8, and 0.9 $\RLo$; $\Tin=$ 5000, 5500, 6000, ..., 10000 K; and $\xi =$ 0.0, -0.1, -0.2, -0.3, -0.5, and -0.75.

For each set of input parameters,  $\mathcal{P}  = (M_1$, $M_2$, $\rin$, $\rout$, $\Tin$, $\xi)$, and  each night's data,  we first minimized the $\chi^2$ contribution of the $g'$ data, $\chig$, against $\Lirr$.  We did not calculate the $i'$ data's contribution to $\chi^2$, $\chii$, simultaneously with the $\chig$-determination because \ELC ignores colour information during simultaneous fits. Thus, once we determined the $\Lirr$ that minimized $\chig$, we calculated the corresponding $i'$ light curve and calculated $\chii$ retaining the $i'-g'$ information. We finally calculated $\chi^2 = \chig + \chii$.

We found that to obtain ``acceptable'' fits (that is, a reduced $\chi^2 \approx 1$), we needed to systematically increase the data's error bars by 0.0164 and 0.0129 mags for the March 08 and March 10 data, respectively. Both of these values are typically less than the 1$\sigma$ errors shown in Figures \ref{fig:glightcurve}.  The error bars shown in Figure \ref{fig:lc_procdata} reflect this increase.

To validate that our spacing of the input parameters is fine enough to effectively probe the $\chi^2$  hyper-surface, we performed additional fits using several test grids of input parameters with spacings at least twice as fine as above.  The best fits obtained in these test calculations were no better than on our standard grid.

\subsection{Possible Influence of the Accretion Stream Hot Spot}
\label{s:hotspot}
The \ELC code does not model contributions from the hot spot where the accretion stream impacts the disk. We argue here that the hot spot's potential contribution can not explain the observed light curves.  However, it may contribute to the disk's overall flux at an important level; if this were the case, then the disk parameters we derive from our fits would be suspect.  Given that  we are unable to meaningfully constrain the parameters of our disk model, as discussed in \S \ref{s:parconst}, this shortcoming does not impact our central results.

The main argument against a modulation of the hot spot's contribution explaining the observed light curve is the orbital phase of flux maximum.  We can compare the time of the ascending node phase, $T_{asc}$, from RXTE timing with our Gemini optical data.  If the sinusoidal modulation represents the heated companion--with light maximum corresponding to donor's superior conjunction--then the $T_{asc}$ from the Gemini data is (TDB) MJD=54169.2784(3). We can compare this with the predicted $T_{asc}$ advanced from the combined orbital solution of \citet{papitto05}, which is (TDB) MJD=54169.280(6), with which it is consistent.  Using the more recent ephemeris of \citet{hartman08} gives a predicted $T_{asc}$ of (TDB) MJD=54169.28000(3), differing from our $T_{asc}$ by 138$\pm$47 seconds, or $1.9\pm0.6$\% of the orbital period.  Either ephemeris confirms that light maximum occurs when the donor is behind the NS.

\citet{hartman08}, which is (TDB) MJD=54169.28000(3).  These times differ by $138\pm47$ seconds, or $2\pm1$\% of the orbital period, confirming light maximum occurs when the donor is behind the NS.

On the other hand, the hot spot maximum should lead the donor's inferior conjunction by some degree. Extrapolating the calculations of \citet{Flannery75} or \citet{Lubow75} to the expected mass ratio range of J1808 predicts that, for a disk extending  only out to the stream's circularization radius, the hot spot maximum should lead the inferior conjunction by a phase of 0.1--0.13. I.e., at a phase 0.37--0.40 later than the observed maximum.  The circularization radius represents the minimum distance the hot spot can be  located from the NS.  The disk likely extends beyond this radius, leading to a hot spot location that leads the donor's inferior conjunction by an even smaller margin. This strongly supports the origin of the modulation as due to the heated secondary, rather than the hot Spot.

\subsection{Parameter Constraints from Light Curve Fits}
\label{s:parconst}
\subsubsection{Model Constraints from Source Distance Estimates}
Before estimating input parameter values from the above $\chi^2$ calculations, we utilized  J1808's distance estimates to only consider theoretical models whose $g'$-band fluxes fall within appropriate ranges.  \citet{galloway06} constrained J1808's distance, $d$, using several different methods.  They modelled the type I X-ray bursts that occurred during the 2002 October outbursts, comparing X-ray burst recurrence times, bolometric fluence, and the ratios of integrated persistent flux to burst fluence. From their best fitting burst models,  they estimate that $d = 3.1$--$3.8$ kpc.  The observed X-ray bursts all showed evidence for radius expansion; for a pure helium NS atmosphere \citep[i.e., consistent with the X-ray bursts being He triggered events,][]{galloway06}, this constrains $d=3.6$ kpc (this estimate decreases if there is any H present in the atmosphere). Finally, based on the fluences and recurrence times of the system's long-term outbursts, \citet{galloway06} estimate a lower limit of $d>3.4$ kpc by assuming the \emph{minimum} time-averaged mass transfer rate onto the NS is set by the binary's gravity-wave driven $\Mdot$ in the conservative limit. 

Based on these estimates, we take as a fiducial constraint $d=3.5$ kpc $\pm$ a $3$\% uncertainty \citep[i.e., the $d=3.4$--$3.6$ kpc of][]{galloway06}.  For comparison, we also carried out our analysis assuming an uncertainty of $\pm 10$\% on $d$.  We then used these $d$ estimates to convert the observed $g'$-values at light curve minima into a $g'$-band luminosity at $\phi=0.0$, $\Lgmin$.  For this we took $A_{g'} = 0.85$ based on $N_H = \ees{1.29}{21}$ cm$^{-2}$ \citep{dickey90}; the corresponding $A_{i'} = 0.47$.  The resulting $\Lgmin$ ranges for the 3\% (10\%) uncertainties on $d$ are $\ees{2.13-2.42 (1.82-2.74)}{30}$ and $\ees{2.47-2.80 (2.11-3.16)}{30}$ erg s$^{-1}$ for the March 08 and March 10 data, respectively.  

\subsubsection{Data and $\Lgmin$-Limit Influences on Input Parameter Constraints: General Trends}
\label{sec:gentrendsparlimits}
There are several general trends in how the input parameters influence the fits. First, the disk parameters ($\rin$, $\rout$, $\Tin$, and $\xi$) are all completely unconstrained by the observations.  Since each of these parameters can influence overall disk brightness and colour, the fact they are not individually constrained is not too surprising. Second, $\chig$ is mainly a function of $\ibin$.  The $g'$-data on its own (see discussion below) rules out $\ibin > 70^\circ$ at the 3-$\sigma$ level (i.e.,  $\Delta \chi^2 = \chi^2>- \chi^2_{\mathrm{min}} > 9.0$ for one free parameter).  Third, the best-fitting $\Lirr$ increases with $\Lgmin$ and, in general,  $\Lirr > 10^{34}$ ergs s$^{-1}$ is required to produce acceptable fits.  

The $\Lgmin$ limits indirectly constrain both $\ibin$ and $M_1$. The relative contribution of the disk vs.\ the donor--which impacts $\chi^2$ primarily through a model's $i'$-band amplitude--impacts both $\Lgmin$ and $M_1$.  Larger $\Lgmin$-values have larger (phase independent) disk contributions, resulting in a lower $i'$-band amplitude.  Larger $M_1$-values have (for a given $\ibin$) larger and hotter donors, which tend to increase $i'$-band amplitudes. These two effects in combination lead to minimum $\chi^2$-values at increasing $\Lgmin$ as the $M_1$-value is increased. Within the $\Lgmin$-range derived from the distance constraints, the minimum $\chi^2$-values generally decrease with increasing $M_1$ while the minima-$\chi^2$ at fixed $M_1$ are increasing functions of $\Lgmin$ (see upper panel of Fig. \ref{fig:deli_lgp}). Thus the lower $\Lgmin$-limit indirectly favours more massive NSs.  On the other hand,  the high-$\Lgmin$ limit indirectly disfavours lower values of $\ibin$ since models with brighter $\Lgmin$ in general require larger $M_2$.

\subsubsection{Input Parameter Constraints}
Taking our fiducial $d$ constraint, we are able to produce acceptable fits to each night's data: the best-fit to the March 08 data had a $\chi^2 = 38.70$ for 37 degrees of freedom (d.o.f), while the best-fit to the March 10 data had a $\chi^2 = 23.94$ (for 17 d.o.f.).
Taken individually, each night's observations place constraints on $\ibin$ and $\Lirr$ at the 2$\sigma$ level.  Both nights require  $\ibin > 30.0$ at a 2$\sigma$ confidence.  For March 08, $\log(\Lirr) = 34.14$--$34.25$ erg s$^{-1}$, while the March 10 limits $\log(\Lirr) = 34.06$--$34.20$ erg s$^{-1}$. The March 08 $i'$ data, since it includes points near both light curve maximum and minimum, constrains our models more strongly then the March 10 data.  Because of this, the March 08 observation, on its own, provides limits on J1808's component masses, requiring  $M_1 \geq 1.8 \msun$ and  $M_2 > 0.07 \msun$ (where these limits are on the two parameters \emph{individually}). The preference for larger $M_1$ results from the need to achieve the observed $i'$-band amplitude given the lower limit on $\Lgmin$ required by the $d$-estimates (see discussion in \S \ref{sec:gentrendsparlimits}).

While the March 10 data set does not by itself provide strong constraints on J1808's binary parameters, fits to it do still favour larger $M_1$ values.  Thus we also considered both nights observations in tandem in order to determine if more stringent constraints on $M_1$ and $M_2$ could be developed.  To do so, for each pair of ($M_1$,$M_2$)-values, we considered all input-parameter sets $\mathcal{P}_k$ (producing a fit with $\chi^2_k$ against March 08 data) and $\mathcal{P}_\ell$ (producing a fit with $\chi^2_\ell$ against the March 10 data) sharing the specified ($M_1$,$M_2$)-values.  As the source distance is treated as a free parameter by \ELC when determining $\chi^2$, we needed to be careful to only consider ($\mathcal{P}_k$, $\mathcal{P}_\ell$)-pairs whose resulting best-fit distance estimates, $\delc$, were mutually consistent between the two nights. We considered a ($\mathcal{P}_k$, $\mathcal{P}_\ell$) to have mutually consistent $\delc$ if $|d_\mathrm{\ELC,i} - d_\mathrm{\ELC,j}| < \Delta d_{3\sigma}$, where $\Delta d_{3\sigma}$ is the maximum change in model distance producing a $\Delta \chi^2 < 9.0$ in a fit's quality.  We determined $\Delta d_{3\sigma}$ by averaging over the required change in $\delc$ for $\approx 60$ of our best-fitting models between both nights.  Finally, for each ($\mathcal{P}_k$, $\mathcal{P}_\ell$) with mutually consistent distances, we calculated  $\chi^2_{k,\ell} = \chi^2_k + \chi^2_\ell$.

\begin{figure}
\plotone{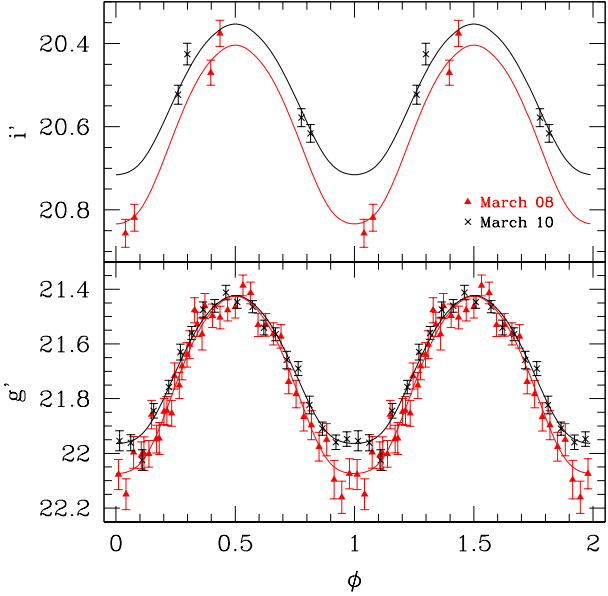}
\caption{The model light curves that, in combination, provide the best fit to both night's data at fixed $M_1$, $M_2$ within our fiducial $\Lgmin$ limits.  The solid lines show the theoretical models.  The symbols show the data (same as in Fig. \ref{fig:lc_procdata}) used to calculate $\chi^2$. }
\label{fig:best_fit_lcs}
\end{figure}

\begin{figure}
\plotone{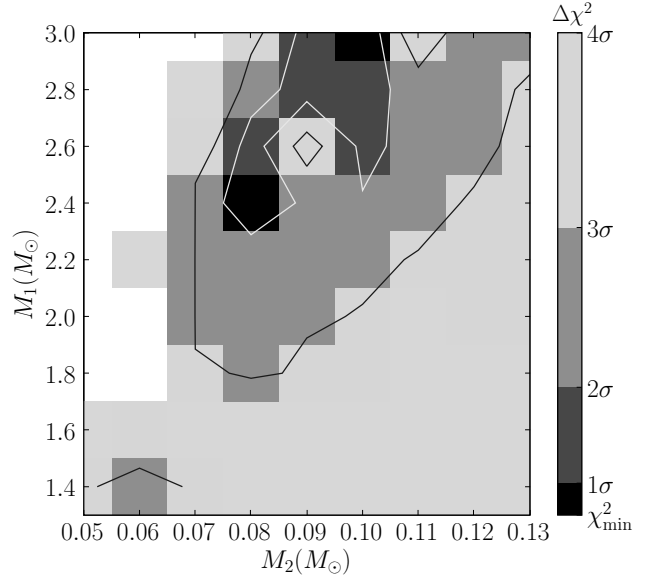}
\caption{Parameter constraints derived by combining fits to both the March 08 and 10 data.  The grey squares show the difference, $\Delta\chi^2_{k,\ell }$, between the minimum value of $\chi^2_{k,\ell}$ at each ($M_1$, $M_2$) and the global $\chi^2_{k,\ell}$-minimum. The greyscale values indicate the uncertainty levels associated with this $\Delta\chi^2_{k,\ell }$ for a single independent parameter (i.e., one and two $\sigma$ corresponds to a $\Delta\chi^2_{k,\ell }= 1.0$ and $4.0$, etc.); thus various regions provide limits on $M_1$ and $M_2$ taken independently.  The solid white and black lines provide approximate contours of the 2$\sigma$ and 3$\sigma$ levels to guide the eye. For this plot, the uncertainty on J1808's distance, $d$, was our fiducial 3\%. }
\label{fig:m1m2_constraints}
\end{figure}

For our fiducial $\Lgmin$ limits, the best fitting summed $\chi^2_{k,\ell} = 63.21$ (for 57 d.o.f.) at $M_1= 3.0 \msun$ and $M_2 = 0.1 \msun$. The light curves of the two models combining to produce this best fit are shown in Figure \ref{fig:best_fit_lcs}.  In Figure \ref{fig:m1m2_constraints} we plot the difference between the minimum $\chi^2_{k,\ell}$ at each  ($M_1$,$M_2$) and this best-fit  $\chi^2_{k,\ell}$-value as grey-scale values keyed to confidence limits.  One can see from  Fig \ref{fig:m1m2_constraints} that our results constrain $M_1 >2.2 \msun$ and $0.07 < M_2 < 0.11 \msun$ at the 2$\sigma$ level; at the same confidence level, $36 < \ibin < 67^\circ$.  Thus, our optical data provides evidence for J1808 harbouring a massive NS, consistent with expectations from the apparent rapid cooling the NS in this system undergoes.

\begin{figure}
\plotone{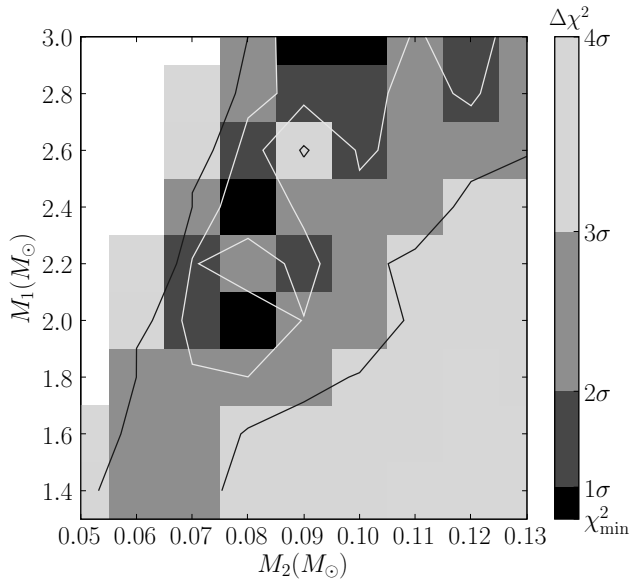}
\caption{Same as Fig. \ref{fig:m1m2_constraints}, but for the case of a 10\% uncertainty on $d$.}
\label{fig:m1m2_constraints_du10}
\end{figure}

The derived lower limit on $M_1$ does depend on the lower $\Lgmin$ limit, which depends on $d$.  Thus, to explore how sensitive our parameter estimates are to the lower $\Lgmin$ limit, we also determined limits on $M_1$ and $M_2$ assuming a 10\% and 20\%  uncertainty on $d$.  The results for the case of a 10\% uncertainty are shown in Figure \ref{fig:m1m2_constraints_du10}. Under these assumptions, we constrain $M_1 > 1.8 \msun$, $M_2 > 0.06 \msun$, and $32 < \ibin < 74^\circ$. For the case of a 20\% $d$-uncertainty, we can not place any limits on  $M_1$ or $M_2$.  Upcoming Gemini time series observations of J1808 will provide better colour information, and may provide stronger evidence for a massive NS.  

\section{Discussion}
\label{s:discussion}
\subsection{Comparison to Prior Work}
There have been several prior observations of SAX J1808 in the optical during quiescence.  
\citet{homer01} reported on observations in both white light and $BVR$ filters.  
Each band's light curve was modulated on the system's $\Porb$ and they estimated the variability in the $V$-band 
to have a semi-amplitude of $\approx 0.08$ mag.  
\citet{campana04} report on $I$-band photometry of J1808, again observing light curve modulation on the system's $\Porb$.
 They also re-evaluated the \citet{homer01} data, reporting a $V$-band modulation semi-amplitude of $0.13 \pm 0.06$ mag.

For our best-fitting models, the $g'$-band modulation semi-amplitude for either night is $\approx 0.3$ mag. 
Since these prior observations did not resolve the fainter star near J1808's position, the fact that we find a 
larger modulation amplitude is not surprising.  Quantitatively, this faint star produces $\approx 66-80$\% the 
$g'$-band flux of J1808 at light curve minimum.  If our observations had not resolved this star, we would have 
found $g'$-band semi-amplitudes of $\approx 0.15 - 0.19$, consistent with the \citet{campana04} analysis of the 
$V$-band data of \citet{homer01}. 

\citet{wang01} used optical observations taken during J1808's 1998 outburst to constrain the system's $\ibin$ and $A_V$.  
For an assumed distance of 3 kpc, their 90\% confidence interval on $i$ is 20-60$^\circ$, with a best-fitting value 
of 37$^\circ$.  Our 2$\sigma$ $\ibin$ limits (36-67$^\circ$) are consistent with these results.

\subsection{Estimates of the Neutron Star's Moment of Inertia}
Since our observations were able to resolve the faint interloping star nearby J1808, it is not surprising that we need a larger $\Lirr$ to explain the larger optical modulation amplitudes we find. Quantitatively, in our good fitting models $\Lirr$ is 
greater by a factor of $\approx 2.5-3$ relative to that found by \citet{campana04}.  As with these prior observations, 
our required $\Lirr$  is 1-2 orders of magnitude greater than the observed quiescent X-ray luminosity $\approx \ees{8}{31}$ erg s$^{-1}$ \citep{heinke08}. We also checked whether the amount of flux required to 
irradiate the donor could be provided by the remnant disk in the system and found that the donor is only able to 
reprocess $\lesssim 1$\% of the disk's luminosity. Thus, this scenario falls 1-2 orders of magnitude short of being 
able to explain the modulation amplitudes.  We conclude that the only source that plausibly can power the optical 
modulation is the spin-down energy of the central NS. As suggested by \citet{burderi03}, this energy is likely tapped 
through a pulsar-wind that turns-on during quiescence (when the mass-transfer rate onto the NS is reduced enough to 
allow the accretion disk to be truncated outside of the NS's light-cylinder).

The luminosity provided by the NS's spin-down is $\dot{E} = 4\pi^2 I \nu \dot{\nu}$, where $I$ is the NS's moment of 
inertia and $\nu$ is its spin-frequency.  \citet{hartman08} have recently measured J1808's long-term 
$\dot{\nu} = -5.6 \pm 2 \times 10^{-16}$ Hz s$^{-1}$.  Thus, the requirement $\Lirr \leq \dot{E}$ allows lower limits to 
be placed on $I$.  To proceed along these lines, we determined lower limits on $\Lirr$ as a function of $M_1$.  These 
limits are derived from the March 08 data due to its larger $g'$-amplitude by taking the minimum March 08 $\Lirr$ within 
the set of all ($\mathcal{P}_k$, $\mathcal{P}_\ell$) with specified $M_1$ and a  $\chi^2_{k,\ell}$ within 3-$\sigma$ of 
the minimum.

For our fiducial 3\% $d$-uncertainties, this minimum $\Lirr \approx \ees{1.5}{34}$ erg s$^{-1}$, almost independent of 
$M_1$.  With $\nu = 400.975$ Hz and $\dot{\nu} = \ees{5.6 \pm 2}{-16}$ Hz s$^{-1}$, this leads to 
$I \gtrsim \ees{1.7 \pm 0.6}{45}$ g cm$^2$ (with the quoted error on $I$ only including the $\dot{\nu}$ uncertainty 
contribution).  The central value of this $I$ limit is certainly consistent with our finding that J1808 harbours a more 
massive NS and, for a range of nuclear equations of state, would require $M_1 > 1.4 \msun$ 
\citep[see Figure 4 of][]{worley08}.  However, the large $\dot{\nu}$ uncertainty prevents our $\Lirr$ limits from 
excluding much of the $I$-$M_1$ parameter space and even $M_1 \approx 1.2 \msun$ are consistent with the lower end of 
the allowed $I$-range.  We note in passing that if the efficiency at which the pulsar's particle wind energy is 
thermalized in the donor's atmosphere were 50\%, then the lower limit on $I$, including the $\dot{\nu}$ uncertainty, 
would rule out $M_1 < 1.8 \msun$ at least for the range of equations of state considered in \citet{worley08}.

\subsection{Neutron Star Cooling Rate and Mass Estimates}
Our $M_1$ estimates can also be connected with the NS cooling rate observed in J1808 \citep[e.g.,][]{heinke07,heinke08}. 
Observations indicate that NSs exhibit a range of cooling rates across both the isolated and accreting sub-populations that are set by the specific neutrino emission processes occurring in the NS core  
\citep[see, e.g.,][]{yakovlev04}. Generally speaking, the most efficient neutrino production process available to lower mass NSs is the modified Urca process, which results in a ``standard'' cooling rate \citep[see, e.g.,][]{yakovlev01}. 
At the higher core densities found in higher mass NSs, several direct Urca processes--depending on the 
constituent nature of matter at high density (i.e., nucleons, hyperons, or ``free'' quarks)--are allowed that provide orders of magnitude faster cooling \citep{yakovlev04} than the standard rate.

The NS mass at which the transition from standard to enhanced cooling occurs and how broad the range of masses over which it occurs depends on the equation of state of matter at supra-nuclear densities and other physics  \citep[such as the occurrence of transitions to various superfluid states and in-medium modifications to the neutrino emission processes][]{yakovlev04,blaschke04}.  Most young isolated NSs appear to cool at rates close to the standard one. On the assumption that most of the NSs in the observed cooling sample have masses consistent with other NS populations, we can infer that those NSs cooling near the standard rate have masses $\approx 1.35 \msun$ \citep{thorsett99}.  Within the isolated NS population there are several clear cases (the Vela pulsar and PSR J0205+64) of objects with cooling rates intermediate between the standard and the fully-developed enhanced rates.  Effective modelling of all the isolated systems strongly indicates that the transition to enhanced cooling has to occur over some mass range and these intermediate cooling-rate systems have masses in this transition interval \citep{yakovlev04,blaschke04}.

Amongst the set of accreting NS systems, there is a similar spread in observed cooling rates.  Again, most systems cool at rates consistent with or slightly faster than the standard rate.  However, J1808 is one of two systems hosting an  accreting NS whose cooling rate is consistent with the fully-developed enhanced rate.  Our constraints limiting $M_1 \gtrsim 1.8 \msun$ are consistent with the picture of high-mass NSs accessing the enhanced cooling mode.  We will postpone a detailed analysis of how J1808's NS cooling rate combined with $M_1$ estimates can constrain matter's equation of state at supra-nuclear densities for a future paper.

We should note that recently \citep{leahy08} have provided estimates for $M_1$ in J1808 based on modelling the pulsed X-ray light curve observed during its 1998 outburst.  Across a set of various models, they find best-fitting $M_1$ values $\lesssim 1.1 \msun$.  Such low $M_1$ values are inconsistent with J1808's enhanced cooling rate given the propensity of data supporting ``normal'' NS masses $\approx 1.35 \msun$ (see above).  The 3$\sigma$ contours quoted for some of the \citet{leahy08} models do include $M_1$ values up to $\approx 1.6 \msun$, which could be consistent with J1808's rapid cooling.  At this same confidence level, our fiducial $M_1$ estimates are barely consistent with those of \citet{leahy08}.  

\subsection{Prior Binary Evolution and Current Donor State}
Given its NS mass function, $\Porb$, and presence of hydrogen \citep{campana04},  J1808 is most likely the descendent of a NS-low mass MS binary whose evolution paralleled that of the related WD accreting cataclysmic variables \citep[see, e.g., ][]{kolb99}. The initial evolution of J1808's LMXB phase was towards shorter $\Porb$ as its then main-sequence (MS) companion was able to maintain thermal equilibrium and contract under mass loss.  As $M_2$ decreased, its Kelvin-Helmholtz time increased, eventually becoming longer than its mass-loss time and the donor began to expand under mass loss, driving the binary to longer $\Porb$.  Standard  evolutionary models find that the resulting $\Porb$ minimum should occur near 70 minutes \citep[while the observed value is closer to 80 minutes][]{kolb99} at which point $M_2 \approx 0.065 \msun$ \citep{politano98,kolb99}.  By the time systems have evolved back out to $\Porb = 2$ hr, $M_2 = 0.02 \msun$ \citep{politano98,deloye08}. As already noted by \citep{bildsten01}, the minimum $M_2$ allowed by the NS's mass function ($0.043 \msun$ for $M_1=1.4 \msun$) is greater than this expected $M_2$-value.  Our 2$\sigma$ limit on the donor's mass, $M_2 > 0.07 \msun$, provides independent evidence for $M_2$ values significantly higher than standard theory predicts for J1808.

At J1808's $\Porb$, if the donor truly has $M_2 > 0.07 \msun$, it would have to have a greater entropy content than the $M_2 \approx 0.06-0.09 \msun$ donors at minimum $\Porb \approx 70-80$ minutes. This possible discrepancy could be explained in at least two ways: (a)  J1808's minimum $\Porb$ was larger than predicted and the MS companion at this point still had sufficient entropy to account for the donor's current state; and/or (b) the donor has been heated during the system's post $\Porb$-minimum evolution. The former could occur if either the donor's KH time was longer than expected (e.g., due to strong external irradiation from the accretion flow) or its mass-loss time was shorter than expected. This could be achieved via an additional angular momentum sink besides gravity-wave radiation losses or to non-conservative mass-transfer effects. Certain models for tidal heating of the donor \citep[e.g.,][]{applegate94} could provide a mechanism for heating the donor during the post $\Porb$-minimum evolution.  We plan to perform detailed modelling of J1808's prior binary evolution to quantify which of these effects can plausibly explain the system's current state. 

\section{Summary}
\label{s:summary}
We undertook optical observations of SAX J1808 in quiescence in $g'$ and $i'$ with Gemini South that were taken nearly coincident with XMM-Newton X-ray observations.  The observations were taken on two separate nights with coverage of slightly more than one of J1808's $\approx 2$ hr orbits each night.  We detected optical modulations in both night's data on the system's $\Porb$ in both bands.  Our observations resolved a nearby interloping star that was unresolved in the observations used in prior light curve analyses \citep{homer01,campana04}.  The optical modulations in our data have a larger amplitude than those measured in these prior efforts, and this amplitude difference is consistent with prior observations being contaminated by the nearby faint star.  Our data also exhibits significant inter-orbit variability.

We modelled the $g'$ and $i'$ light curves obtained using the light curve modelling program \ELC.  Our model's free parameters were the NS and donor masses ($M_1$, $M_2$), parameters describing the disk ($\rin$, $\rout$, $\Tin$, and $\xi$), and the luminosity irradiating the donor, $\Lirr$, that powers the optical modulation.  The binary inclination, $\ibin$, and orbital separation are then determined from the measured NS mass function and system $\Porb$.  To fit our data, we constructed a grid over  ($M_1$, $M_2$, $\Tin$, $\rin$, $\rout$, $\xi$) and for each vertex of this grid, optimized $\Lirr$, performing the most extensive modelling of J1808's optical behaviour to date.

We derived estimates for our input parameters from the subset of our models whose $g'$ luminosity was consistent with the optical data given J1808's distance, $d$, constraints. At the 2$\sigma$ confidence level, we constrain $\ibin$ = 36--67$^\circ$.  We also constrain $M_1$ and $M_2$, but these limits are sensitive to the uncertainties in $d$ (see the Appendix).  For our fiducial 3\% $d$-uncertainty \citep{galloway06}, we constrain $M_1 > 2.2 \msun$ and $0.07 < M_2 < 0.11 \msun$.  For a 10\% d-uncertainty, these limits expand to $M_1 > 1.8$ and $M_2 > 0.06 \msun$. We treat all the disk parameters as independent (since a quiescent disk is not in a steady state), and this freedom prevented us from deriving any constraints on the disk's structure.

To power the optical modulation, $\Lirr = \ees{1.15-1.78}{34}$ erg s$^{-1}$.  The almost coincident X-ray observation determined the X-ray luminosity of the source at this epoch to be $\ees{7.9 \pm 0.7}{31}$ erg s$^{-1}$ \citep{heinke08}, far below that required to power the optical variability.  However, the measured NS spin-down rate \citep{hartman08} can reasonably provide the required $\Lirr$.  As suggested previously \citep{burderi03}, this could indicate that radio pulsar activity switches on in J1808 during its quiescent periods and the resulting pulsar particle wind provides the necessary flux to illuminate the donor's day-time side.  If the donor is less than 100\% efficient in converting this particle flux to thermal radiation, then the required $\Lirr$ increases above our quoted range. An efficiency of 50\% would require a sufficiently large NS moment of inertia that $M_1 < 1.8 \msun$ would be ruled out for many reasonable nuclear equations of state.

Our estimates of $M_1 > 1.8 \msun$ are consistent with J1808's very low thermal X-ray luminosity \citep{heinke08}.  Such massive NSs are  expected to cool rapidly due to direct Urca neutrino emission processes that can not occur in lower mass NSs. Our estimates that $M_2 > 0.07 \msun$ also provide independent support (in addition to the NS's mass function) that the donor in this system has a significantly higher entropy than expected from standard binary evolution models. This would indicate that any of several processes that could increase the system's minimum $\Porb$ or heat the donor have been operative in the system.

\section*{Acknowledgments}
CJD thanks J. Orosz for providing the \ELC code and for assistance in using this code. COH thanks A. Bonanos for suggesting the use of ISIS and providing assistance with using it, and D. Chakrabarty for calling our attention to the existence of a star close to J1808. We thank the anonymous referee for pointing out citation omissions. COH acknowledges support from the Lindheimer Postdoctoral Fellowship at Northwestern University, NASA {\it Chandra} grants G07-8078X, G08-9053X, and G08-9085X, and NASA XMM grant NNX06AH62G.  CJD acknowledges support from NASA \textit{Chandra} grant TM7-8007X and XMM grant NNX06AH62G. Support for this work is provided in part by the Theoretical Institute for Advanced Research in Astrophysics (TIARA) operated under Academia Sinica and the National Science Council Excellence Projects program in Taiwan administered through grant number NSC 96-2752-M-007-007-PAE. PGJ acknowledges support from the Netherlands Organisation for Scientific Research. Based on observations obtained at the Gemini Observatory, which is operated by the Association of Universities for Research in Astronomy, Inc., under a cooperative agreement with the NSF on behalf of the Gemini partnership: The National Science Foundation (United States), the Particle Physics and Astronomy Research Council (United Kingdom), the National Research Council (Canada), CONICYT (Chile), the Australian Research Council (Australia), CNPq (Brazil), and CONICET (Argentina).

\appendix
\section{Further Details on how the Data and Distance Constraints Determine Input Parameters}
\begin{figure}
\plotone{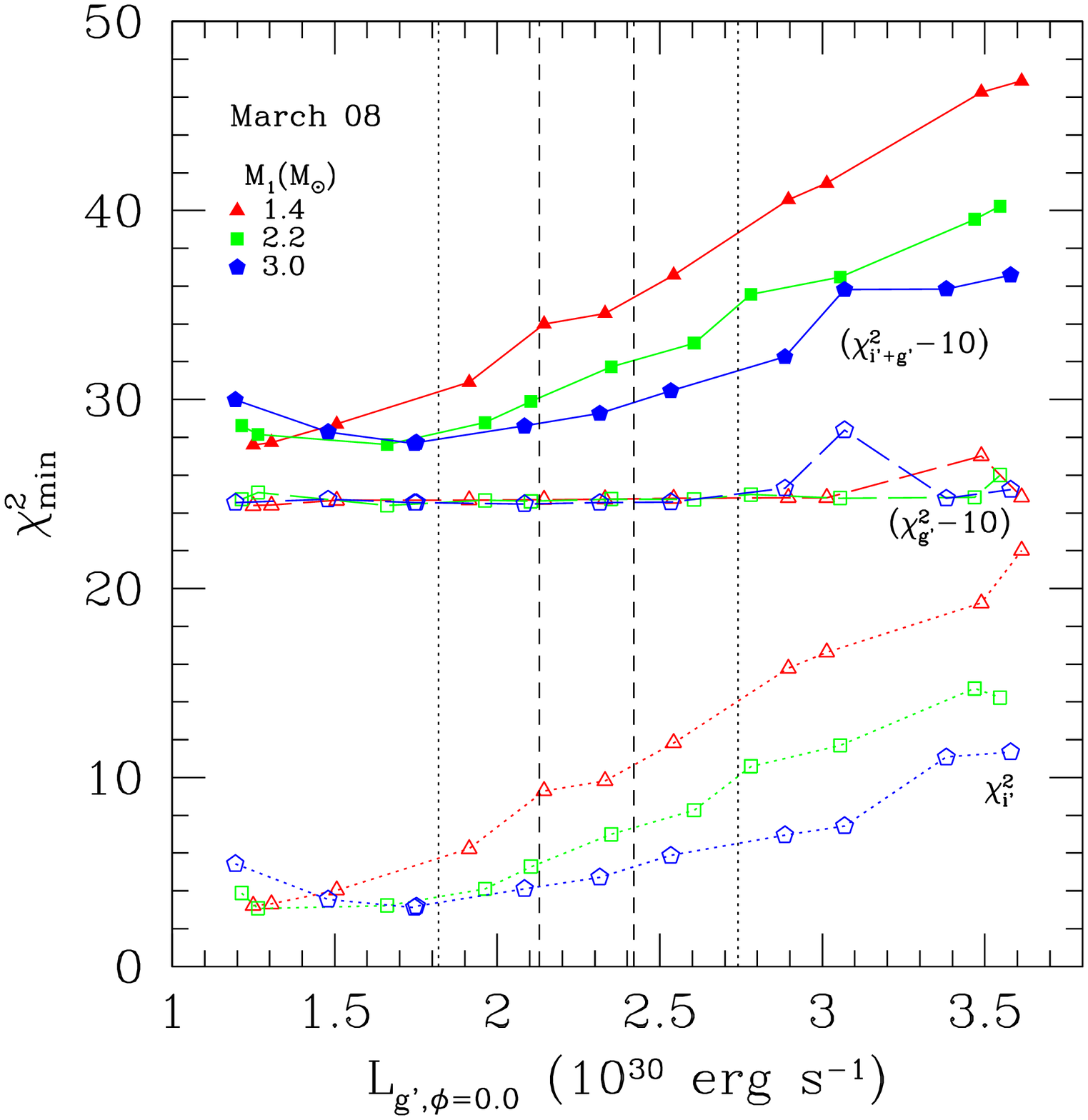}
\caption{The evolution of minimum $\chi^2$ at fixed $M_1$ versus $\Lgmin$ for the March 08 data.  
Solid lines show the overall $\chi^2$-minima as a function of $\Lgmin$, while the dashed and dotted coloured lines 
show the $g'$ and $i'$ data's' contributions to the overall $\chi^2$.  The vertical dashed and dotted lines 
provide the $\Lgmin$ limits on the March 08 data assuming a 3\% and 10\% uncertainty on $d$.}
\label{fig:chi2vLg}
\end{figure}

 Here we provide some of the background details concerning how the data and $\Lgmin$ limits contribute to constraints on input parameters.  Figure \ref{fig:chi2vLg} shows, for the March 08 data, how the minimum of $\chi^2$ at
 specified $M_1$ varies as a function of $\Lgmin$.  The solid coloured lines show the total $\chi^2$, while the 
coloured dashed and dotted lines show the $g'$ and $i'$ data's contributions to the total $\chi^2$ (both the $g'$ 
and total values are decreased by 10.0 to reduce the vertical scale of the plot).  Due to the fact that we minimize 
$\chi^2_{g'}$ against $\Lirr$ first and then determine $\chi^2_{i'}$ afterwards, the $g'$ data's amplitude is 
relatively well fit across all models and there is little variation in $\chi^2_{g'}$ with $\Lgmin$ or $M_1$.

All the differences in $\chi^2$ with $\Lgmin$ therefore are produced by the $i'$ data.  For each $M_1$, $\chi^2_{i'}$ 
has a minimum at different values of $\Lgmin$, with the $\Lgmin$ at these minima increasing with $M_1$.  
Above $\Lgmin > \ees{2}{30}$ ergs s$^{-1}$, larger $M_1$ values produce better fits to the $i'$ data.  
The vertical dashed (dotted) lines show the March 08 limits on $\Lgmin$ for our 3\% (10\%) uncertainties on $d$.  
We see that increasing the lower limit of $\Lgmin$ removes preferentially the best fitting models at lower $M_1$.  
This is the origin of the distance limit producing a preference for a massive $M_1$.

\begin{figure}
\plotone{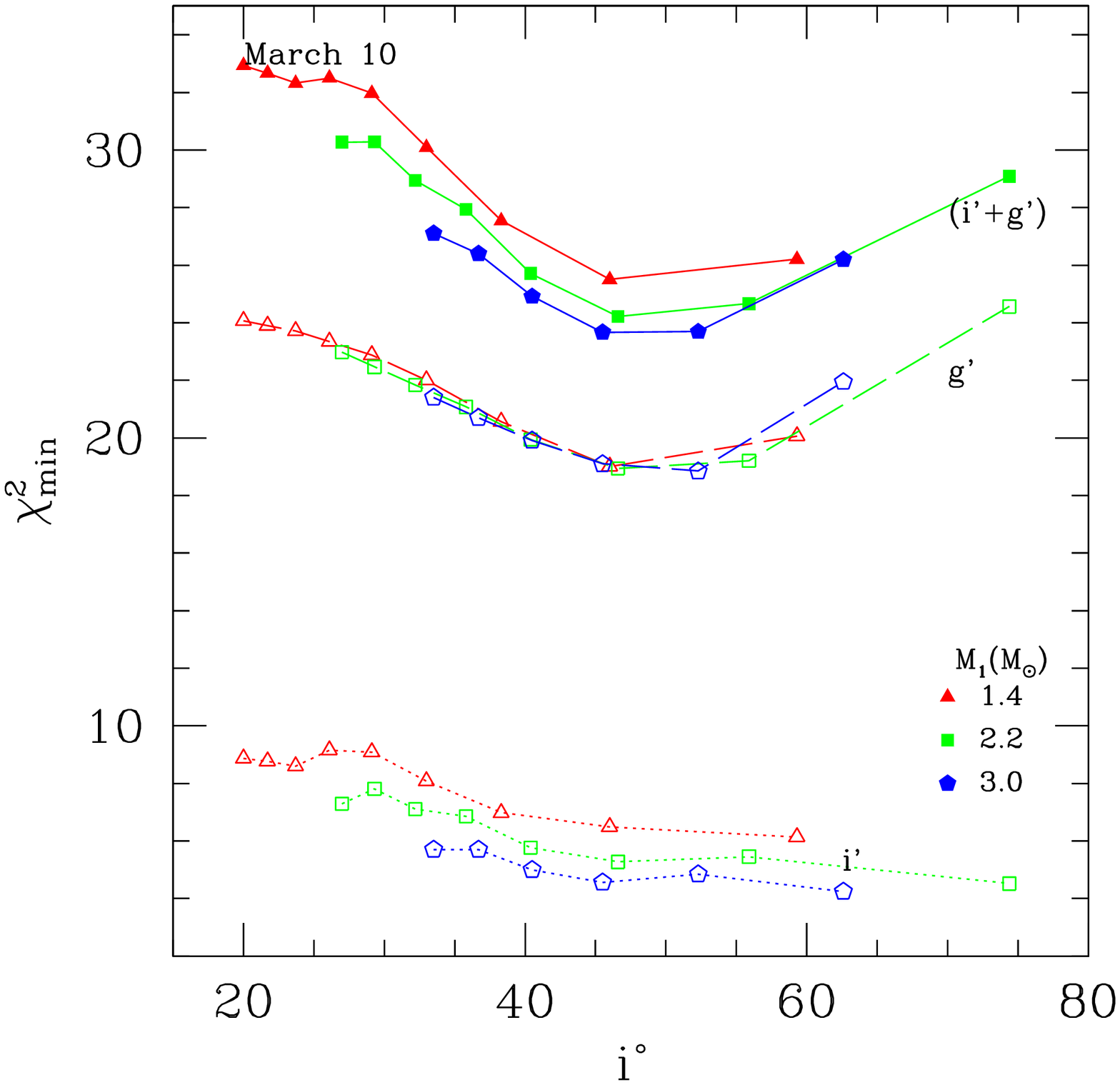}
\caption{The variation of minimum $\chi^2$ vs. $\ibin$ at fixed values of $M_1$. 
 Line colours and styles have the same meaning as in Fig. \ref{fig:chi2vLg}.}
\label{fig:chi2vi}
\end{figure}

Before going into more detail about why larger $M_1$ values lead to better fits in the $\Lgmin$ range, we'll discuss how 
the data sets the $\ibin$ constraints.  In Figure \ref{fig:chi2vi}, we show the minimum $\chi^2$ versus $\ibin$, again 
at fixed values of $M_1$ for the March 10 data.  Colours and line styles in this plot have the same meaning as in 
Fig. \ref{fig:chi2vLg}, and the data only include those systems satisfying the $\Lgmin$ constraints. The $g'$ data in 
this case, since we have a full light curve, on its own constrains the orbital inclination to a reasonable degree.  
The $i'$ data, on the other hand for the most part provides a weaker trend versus $\ibin$, except 
at $\ibin \lesssim 30^\circ$.  The increasing $\chi^2_{i'}$ at the lower $\ibin$ range is actually contributed to by 
the $\Lgmin$ upper-limits. Higher $\Lgmin$ values tend to have better fits with more massive $M_2$ values.  Removing the 
brighter systems from the sample tends to remove the best fitting models at lower $\ibin$.   Thus in combination, 
the $g'$ light curve morphology and the upper-limits on $\Lgmin$ favour $\ibin$ in the range $30 < \ibin < 70^\circ$.   
Unlike the case for $\Lgmin$, there is no apparent difference in best-fitting $\ibin$ between $M_1$ values.

\begin{figure}
\plotone{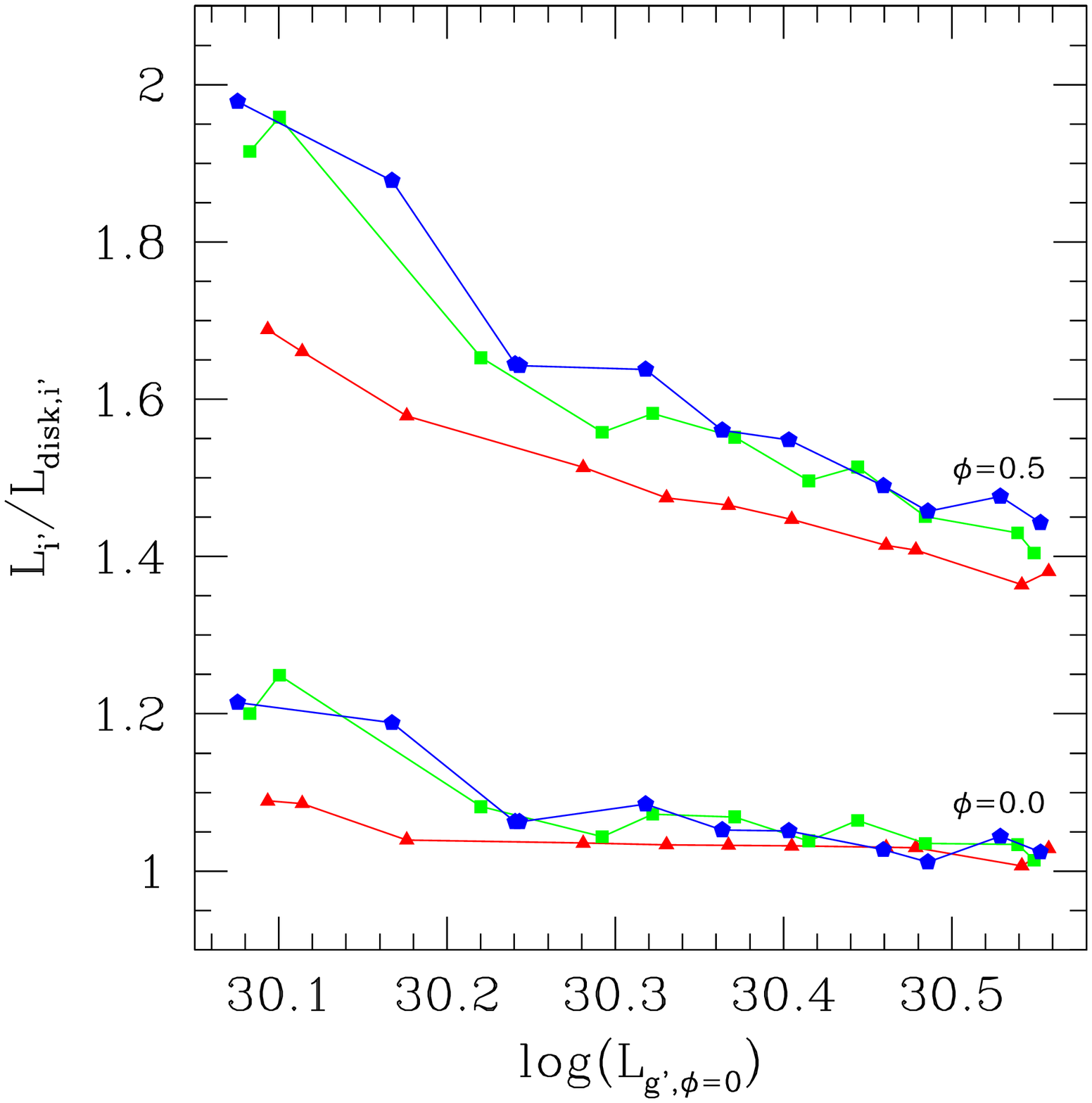}
\caption{The ratio of total $i'$ luminosity to that contributed by the disk alone as a function of 
$\Lgmin$ at $\phi=0.0$ and $0.0$ and fixed $M_1$.  The different colours indicate the same $M_1$ values as in 
the previous two plots.}
\label{fig:Liminmax}
\end{figure}

The origin of the $i'$ data preferring heavy NSs lies in the amplitude of $i'$ light curves.  In particular, 
since the March 08 $i'$-data probes almost the full amplitude of the light curve, stronger constraints on input 
parameters are derived from it versus the March 10 data.  Why is the $i'$ amplitude important?  
Figure \ref{fig:Liminmax} shows the ratio of total $i'$ luminosity to that contributed by the disk at both $\phi=0.0$ 
and $0.50$ as a function of $\Lgmin$ (this is for the best fitting models at fixed $M_1$ with the same values as in the 
previous two figures).  As $\Lgmin$ increases, the disk contribution to the total light increases, but is typically less 
in the systems with more massive $M_1$.  The evolution in disk contribution at $\phi=0.5$ is particularly strong.  

With a strongly varying disk contribution to the total light at $\phi=0.5$, the overall $i'$ amplitude will vary with 
$\Lgmin$ and this amplitude differs between $M_1$ values.  Typically, the best-fitting models at higher $M_1$ 
and fixed $\Lgmin$ have a larger $i'$ amplitude.  This is due to heavier $M_1$ typically having more massive $M_2$ 
which are \emph{intrinsically} brighter (this can be seen by the difference even at $\phi=0.0$, where larger $M_1$ 
systems have smaller disk contributions). 

\begin{figure}
\plotone{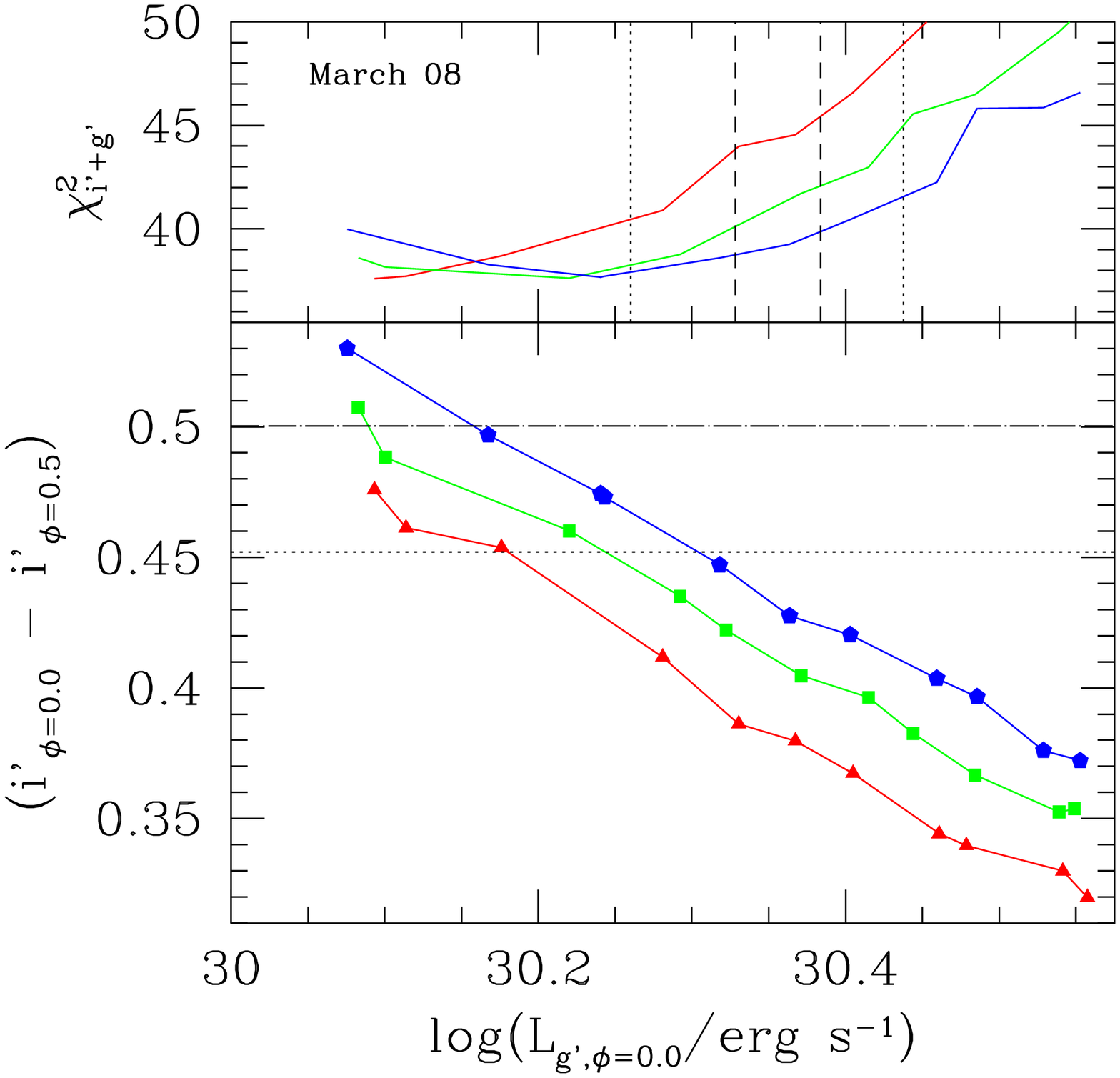}
\caption{Lower panel: the amplitude in $i'$ produced by the models in Fig. \ref{fig:Liminmax} (coloured lines) versus the 
March 08 $i'$ data's amplitude (dashed line with 1-$\sigma$ error bar given by dotted line).  Upper panel: the $\chi^2$
 of these models with the $\Lgmin$ limits corresponding to a 3 and 10\% uncertainty on the $d$ value (vertical dashed 
and dotted lines, respectively). }
\label{fig:deli_lgp}
\end{figure}

The lower panel of figure \ref{fig:deli_lgp} looks at how the $i'$ amplitude varies with $\Lgmin$ and $M_1$ for the 
best fitting models shown in Fig. \ref{fig:Liminmax}.  It can be seen clearly there that larger $M_1$ values produce 
larger $i'$ amplitudes at fixed $\Lgmin$.  The horizontal dashed and dotted lines show the approximate amplitude and 
lower 1$\sigma$ error bar for the March 08 $i'$ data.  The fact that the best-fitting models with larger $M_1$ agree 
in $i'$ amplitude with the data at larger $\Lgmin$ values is the ultimate reason why the $\Lgmin$ limits favour 
massive $M_1$.  This is shown in the upper panel, where the minimum $\chi^2$ for the three different $M_1$ values is 
reproduced along with the $\Lgmin$ limits corresponding to the 3 and 10\% $d$ uncertainties.  This also shows why if 
$d$ were only known to 20\%, there are no constraints on $M_1$: the best fitting low-$M_1$ models at lower $\Lgmin$ 
are then not excluded.  Thus, the lower $\Lgmin$ limit, given our current data, is critical to the constraints on
 $M_1$ derived here.

\end{document}